\documentclass[a4paper, 12pt]{article}      
\usepackage{a4}
\usepackage{amsfonts}
\usepackage{amssymb}
\usepackage{epsfig}

\author{}

\newcommand{\be}{\begin{equation}}
\newcommand{\ee}{\end{equation}}
\newcommand{\ba}{\begin{array}}
\newcommand{\ea}{\end{array}}
\newcommand{\bea}{\begin{eqnarray}}
\newcommand{\eea}{\end{eqnarray}}

\def\IR{\relax{\rm I\kern-.18em R}}

\def\IP{\relax{\rm I\kern-.18em P}}
\def\inbar{\vrule height1.5ex width.4pt depth0pt}
\def\IC{\relax\,\hbox{$\inbar\kern-.3em{\rm C}$}}

\def\K3{{\bf K3}}

\def\n2d{\cN_{V^*}^{\otimes 2}}

\def\IC{\mathbb{C}}

\def\IR{\mathbb{R}}

\def\IP{\mathbb{P}}

\def\cN{{\mathcal N}}

\begin{document}

\title{
\begin{flushright} \vspace{-4cm}
{\small MPP-2010-131 \\
 \small LMU-ASC 75/10\\}
\end{flushright}
\vspace{2.5cm}
 T-duality and closed string non-commutative (doubled) geometry} 

\date{}

\maketitle

\vspace{-1.4cm}


\begin{center}

{\bf Dieter L\"ust}$^{a,b,}$\footnote{dieter.luest@lmu.de, luest@mppmu.mpg.de}

\vspace{.6truecm}

{\em $^a$Arnold Sommerfeld Center for Theoretical Physics\\
Department f\"ur Physik, Ludwig-Maximilians-Universit\"at M\"unchen\\
Theresienstr.~37, 80333 M\"unchen, Germany}

\vspace{.4truecm}

{\em $^b$Max-Planck-Institut f\"ur Physik\\
F\"ohringer Ring 6, 80805 M\"unchen, Germany}

\end{center}

\vspace{0.5cm}

\begin{abstract}
\noindent  
 We provide some evidence that  closed string coordinates will become non-commuta\-tive
 turning on geometrical fluxes and/or $H$-field flux background in closed string compactifications. This is in analogy to open
 string non-commutativity on the world volume of D-branes with $B$- and $F$-field background.
 The class of 3-dimensional backgrounds we are studying are twisted tori (fibrations
 of a 2-torus over a circle) and the their T-dual $H$-field, 3-form flux backgrounds (T-folds).
 The spatial non-commutativity arises due to the non-trivial monodromies of the toroidal 
 K\"ahler resp. complex structure moduli fields, when going around the closed string along the circle direction.
  In addition we study
 closed string non-commutativity in the context of doubled
 geometry, where we argue that in general a non-commutative closed string background is T-dual
 to a commutative closed string background and vice versa. We also discuss the corresponding spatial uncertainty relations.
 Finally, in analogy to open string boundary conditions, we also argue that closed string momentum and winding modes
 define in some sense D-branes in closed string doubled geometry.
  \end{abstract}

\thispagestyle{empty}       
\clearpage

\section{Introduction}

T-duality is one of the most interesting symmetries in string theory.  Most importantly, T-duality is a stringy symmetry in the sense
that it arises due to the extended nature of the string in contrast to a point particle.
In its simplest and also most transparent manifestation, T-duality
arises when one
compactifies the closed string on a D-dimensional torus (for a review see \cite{Giveon:1994fu}). 
In case of compactification on a one-dimensional circle $S^1$ of radius $R$, T-duality
is just the invariance of the string theory under the inversion of the radius
\begin{equation}
T:\quad R\, \, \longleftrightarrow\, \, {\alpha'\over R}\, .\label{Tduality}
\end{equation}
In fact, this symmetry acts on the string spectrum exchanging momentum modes with
momenta $p=M/R$ by closed string winding modes of dual momenta $\tilde p=(\alpha')^{-1}NR$, i.e.
T-duality acts on the integer momentum and winding numbers simply as
\begin{equation}
T:\quad M\, \, \longleftrightarrow\, \, N\, .\label{Tduality1}
\end{equation}
Since this symmetry has a self-dual radius,
\begin{equation}
R_c=\sqrt{\alpha'}=l_s\,,
\end{equation}
it follows that the stringy moduli space of in-equivalent circle compactifications is given by the
interval 
\begin{equation}
R\geq R_c\, .
\end{equation}
In other words, a compact space of size $R>R_c$ is
completely indistinguishable from a compact space with size $R<R_c$, and hence $R_c$ 
is the shortest possible radius that characterizes the closed string compactification on a circle.

The emergence of a minimal  distance in string compactification means that the notion
of classical geometry breaks down at distances around $R\sim R_c$.
This concept of a minimal observable distance in string theory arises also in a different context,
namely in the context of high energy string scattering \cite{Amati:1987wq}.
Here, it was argued that string high energy scattering experiments leads to a modification
of Heisenberg's uncertainty relation of the following form:
\begin{equation}\label{modheis}
\Delta x\simeq {1\over \Delta p}+{\alpha'\Delta p}\, .
\end{equation}
It again follows that with a  string scattering experiment one cannot resolve distances smaller $\Delta x_{\rm min}=\sqrt{\alpha'}$, i.e.
\begin{equation}
\Delta x\geq\sqrt{\alpha' }\, .
\end{equation}

So far, closed string non-commutativity with respect to the space coordinates was not much discussed in the literature
(see e.g. \cite{Frohlich:1993es,DeRisi:2002gt}).
Non-commutative string geometry was mainly established for open strings   \cite{Witten:1985cc,nappi,Chu:1998qz,Schomerus:1999ug,seiwit}.
Specifically it was shown that for open strings with mixed Dirichlet-Neumann boundary conditions on  Dp-branes
with non-vanishing $B$-field or gauge field $F$-background,
the open string coordinates at the two ends of the open string, i.e. at $\sigma=0,\pi$, are non-commutative:
\begin{equation}
[X_1(\tau,\sigma),X_2(\tau,\sigma')]_{\sigma,\sigma'=0,\pi}=i\Theta\, .
\end{equation}
The deformation parameter $\Theta$ is determined by the open string background parameter ${\cal F}_{ij}=B_{ij}+F_{ij}=\epsilon_{ij}{\cal F}$
in the following way:
\begin{equation}
 \Theta={2\pi\alpha'{\cal F}\over 1+{\cal F}^2}\,
\end{equation}
Performing a T-duality transformation, the mixed D-N boundary conditions become purely Neumann,
and the Dp-branes with ${\cal F}$-flux become D(p-1)-branes that intersect at a certain angle $\theta$, where $\cot\theta={\cal F}$ \cite{angles}.
Now after T-duality the dual open string coordinates are fully commutative.\footnote{If
one wants to construct a full fledged open string compactification on a non-commutative torus resp. on the dual torus with intersecting
 D-branes,
further conditions like the vanishing of 1-loop open string tadpoles have to be satisfied \cite{Blumenhagen:2000wh}.}

However, also for an extended object like a closed string, T-duality tells us that there is a minimal length that can be dissolved by the string, 
and hence space 
should be come fuzzy at the string scale. Therefore for extended closed strings with non-trivial winding numbers
one could expect a space non-commutativity of the form
\begin{equation}
[X,Y]\neq 0\, ,
\end{equation}
corresponding to a spatial uncertainty relation
\begin{equation}
\Delta X \Delta Y\neq 0\, .
\end{equation}
On the other hand, a pure momentum state of the form $\exp (p X)$ is still a point  like object. Hence space coordinates  should be still commutative for a pure momentum spaces.
Using T-duality, a momentum state becomes a winding state in the dual space with dual coordinates $\tilde X$, whereas  states, which are originally winding modes,
become  momentum states in the dual geometry. On the basis of this behavior, we could deduce that momentum states possibly see
a non-commutative dual space with
\begin{equation}
[\tilde X,\tilde Y]\neq 0\, ,
\end{equation}
but for winding states the dual geometry is commutative.
So again, like for the open string, T-duality exchanges commutative with non-commutative closed string coordinates.
Hence if by some reason the $\tilde X$-space is deformed to be non-commutative, it almost trivially follows that in the T-dual geometry
the $X$-space becomes non-commutative, and vice versa. 

In this paper we show that the natural framework to discuss closed string non-commutative geometry is indeed T-duality in close relation with doubled geometry \cite{Hull:2006va,Hull:2007jy,Hull:2009sg}
resp. with doubled field theory \cite{Hull:2009mi,Hull:2009zb,Hohm:2010jy,Hohm:2010pp};  here the usual closed string
position space, denoted by string coordinates $X^i$, is enlarged to contain  in addition also  the dual closed
string coordinates $\tilde X^i$. We will give some evidence that for certain geometrical backgrounds, namely for geometric
flux compactifications on twisted tori 
\cite{Hellerman:2002ax,Dabholkar:2002sy,Kachru:2002sk,Derendinger:2004jn,Hull:2005hk,Shelton:2005cf,Dabholkar:2005ve,Grana:2006kf}, 
space with coordinates  becomes non-commutative. In fact, these backgrounds are in general T-dual to non-geometric compactifications with a non-trivial a $H$-field background. 
The main result of this paper will be  to give  evidence that a  closed string on a  curved space (3-dimensional torus) with a non-trivial
geometrical flux and/or  $H$-flux sees
a non-commu\-ta\-tive geometry  along two  directions, which is determined by the (dual) momentum $p^3$ in the remaining third  direction:
\begin{equation}
\lbrack X^1, X^2\rbrack \sim 2i(l_s)^3\,p^3\, .
\end{equation}
Therefore for non-vansihing (dual) momenta  in the third direction, there will be a  spatial uncertainty in the other two directions.

In the appendix we will briefly review the closed string boundary conditions in doubled $(X,\tilde X)$-geometry and
the action of T-duality. Momentum resp. winding closed strings are analogous to open strings with Neumann resp. Dirichlet bound
conditions; hence we will argue that the choice of closed string boundary conditions also defines a kind of brane in the doubled geometry.

\section{Closed string doubled geometry, T-duality and non-commutativity}

In this section we recall some known and basic facts about closed strings moving on a compact torus, where
we will add a few more remarks about the analogy between open and closed strings in the appendix.  First consider
the most simple case of a closed string moving on a circle with radius $R$. Its most general mode expansion has the form
\begin{equation}
X(\tau,\sigma)=X_L(\tau+\sigma)+X_R(\tau-\sigma)\, ,
\end{equation}
\bea
X_L(\tau+\sigma)&=&{x\over 2}+p_L(\tau+\sigma)+i\sqrt{{\alpha'\over 2}}\sum_{n\neq 0}{1\over n}\alpha_ne^{-in(\tau+\sigma)}\, ,\nonumber\\
X_R(\tau-\sigma)&=&{x\over 2}+p_R(\tau-\sigma)+i\sqrt{{\alpha'\over 2}}\sum_{n\neq 0}{1\over n}\tilde\alpha_ne^{-in(\tau-\sigma)}\, ,
\eea
where the left- and right-moving momenta are defined as
\bea
p_L&=&{1\over 2}\biggl({M\over R}+(\alpha')^{-1}NR\biggr)\, ,\nonumber\\
p_R&=&{1\over 2}\biggl({M\over R}-(\alpha')^{-1}NR\biggr)\, .
\eea
Here, the integers $M$ are the Kaluza-Klein momenta, whereas the integers $N$ denote the winding numbers of the
closed string around the circle $S^1$.
The quantized momentum variables $p$ and the dual momenta $\tilde p$ are then given as
\begin{equation}
p=p_L+p_R={M\over R}\, ,\quad \tilde p=p_L-p_R=(\alpha')^{-1}NR\, .
\end{equation}
T-duality (see eqs.(\ref{Tduality}) and (\ref{Tduality1})) acts in the well-known fashion on the momenta:
\begin{equation}
T:\quad p\, \, \longleftrightarrow\, \,\,\, \tilde p\, ,\quad p_L\longleftrightarrow p_L, ,\quad p_R\longleftrightarrow -p_R\, .
\end{equation}

Just as for the canonical momenta, we can also introduce a dual space coordinate $\tilde X(\tau,\sigma)=X_L-X_R$.  The pair $(X,\tilde X)$
are the coordinates of so-called doubled geometry.
So finally, T-duality acts on the position variables as
\begin{equation}
T:\quad X\, \, \longleftrightarrow\, \, \tilde X\, ,\quad X_L\longleftrightarrow X_L, ,\quad X_R\longleftrightarrow -X_R\, .
\end{equation}

All this can easily be generalized to the compactification of the closed string on a d-dimensional
torus $T^d$. Here we considering a constant metric background $G_{ij}$ as well as a constant background for
the antisymmetric tensor field $B_{ij}$.
The canonical left- and right-moving momenta contain also the B-field background and have the following form:
\begin{equation}\label{canmom}
    p^i_{L,R} = {\alpha'\over2} \left(G_{ij}M^j \pm  \frac{1}{\alpha'}(G^{ij} \mp B^{ij})\, N_j\right) 
  \end{equation}
Then the T-duality symmetries are given by $SO(d,d;{\mathbb Z})$ transformations.

For non-constant backgrounds, one can also apply T-duality in the $x$-direction, assuming that the background does not depend on
the $x$-coordinate. Then using the Buscher rules \cite{Buscher:1987qj,Buscher:1987sk},
T-duality in the $x$ direction
provides the following new background:
\begin{eqnarray}
\tilde G_{xx} &=& \frac{1}{G_{xx}}, \qquad
\tilde G_{x i}  = -\frac{B_{x i}}{G_{xx}}, \qquad
\tilde B_{x i} = -\frac{G_{x i}}{G_{xx}} \\
\tilde G_{i j}  &=&  G_{i j} - \frac{G_{x i}G_{x j} - B_{x i} B_{x j}}{G_{xx} } \\
\tilde B_{ij}  &=&  B_{ij} - \frac{G_{x i}B_{x j} - B_{x i} G_{x j}}{G_{xx} } \\
e^{\tilde \phi} &=& \frac{e^{\phi}}{\sqrt{G_{xx}}}
\label{buscherrules}
\end{eqnarray}
This basically means that a flux background with non-vanishing $B_{xi}$ is T-dualized into
a purely geometric back ground with off-diagonal metric $G_{xi}$ and vice versa.
Switching from $B$ to $H$, a non-vanishing $H_{xyz}$ gets T-dualized along the $x$
direction in a metric background, which we call $G_{yz}=f^x_{yz}$
\begin{equation}\label{geomh}
H_{xyz} \stackrel{T_x}{\longrightarrow} f^x_{yz}.
\end{equation}
Since these metric components arise from the $H$-flux after T-duality, one often calls
the $f^x_{yz}$'s  metric or also geometric fluxes. As we will discuss in the next chapter, they often appear
in twisted tori compactifications
where they correspond to a certain, underlying algebraic
structure.

In the following we are mostly  interested in the case $d=2$, the compactification on a two-dimensional torus $T^2$,
which is
defined by the two vectors $e_1=R_1$ and $e_2=R_2e^{i\alpha}$, and
with additional
$B$-field background $B_{ij}=\epsilon_{ij}B$. 
This background is then conveniently characterized by two complex parameters: first the complex structure of the torus,
denoted by $\tau$, and second by the complexified K\"ahler parameter, denoted by $\rho$:
\bea
\tau&=&{e_2\over e_1}={R_2\over R_1}e^{i\alpha}\, ,\nonumber\\
\rho&=&B+iR_1R_2\sin\alpha\, .
\eea
The full T-duality group is given by all possible
 $SO(2,2;{\mathbb Z})$ transformations. First, it contains the global diffeomorphisms of $T^2$ given by $SL(2,{\mathbb Z})_\tau$ modular transformations 
 \begin{equation}
 \tau\rightarrow{a\tau+b\over c\tau +d}\, .
 \end{equation}
 Second the discrete shifts in $B$, $B\rightarrow B+n$, together with the overall T-duality transformation, $\rho\rightarrow-1/\rho$, in the $x_1$- and in $x_2$-directions
 generate the
 target space modular group $SL(2,{\mathbb Z})_\rho$ acting as
 \begin{equation}
 \rho\rightarrow{a\rho+b\over c\rho +d}\, .
 \end{equation}
 Finally, the mirror symmetry, i.e. the T-duality in $x_1$-direction is nothing else then the exchange
 \begin{equation}
 \tau\leftrightarrow\rho\, .
 \end{equation}
 Clearly this transformation exchanges the $B$-field with $\Re\tau$, which is just proportional to the off-diagonal metric component.

It is now also straightforward to work out the action of the various target space duality transformations on the two-dimensional string coordinates
of $T^2$. Introducing complex coordinates 
\bea
X=X^1+iX^2=X_L+X_R\, ,
 \eea with $X_L=X^1_L+iX^2_L$ and likewise for $X_R$, we see  that the  geometric
$SL(2,{\mathbb Z})_\tau$ transformations  act as left-right symmetric  transformations on the complex coordinates. E.g. one obtains that
\bea\label{symrot}
\tau\rightarrow -1/\tau\, :\qquad &{~}&X_L\rightarrow e^{i\theta}X_L\, ,\quad \theta=-\pi/ 2\, ,\nonumber\\
&{~}&X_R\rightarrow e^{i\theta}X_R\, .
\eea
This is nothing else than a ${\mathbb Z}_4$ transformation on the coordinates:
\begin{equation}
X_1\rightarrow X_2, \qquad X_2\rightarrow -X_1\, .
\end{equation}
Other elements of $SL(2,{\mathbb Z})_\tau$ like the order three transformation $\tau\rightarrow -1/(\tau+1)$ act as symmetric ${\mathbb Z}_6$ rotations 
on the left- and right-moving coordinates.

On the other hand, the target space duality transformations in $SL(2,{\mathbb Z})_\rho$ basically act as asymmetric rotations on 
$X_L$ and $X_R$ \cite{Ibanez:1990ju}. 
E.g. consider the 
overall duality transformation $\rho\rightarrow -1/\rho$. It acts as follows:
\bea\label{asymrot}
\rho\rightarrow -1/\rho\, :\qquad &{~}&X_L\rightarrow e^{i\theta}X_L\, ,\quad \theta=-\pi/ 2\, ,\nonumber\\
&{~}&X_R\rightarrow e^{-i\theta}X_R\, .\nonumber \\
\eea

Finally let us also consider the dual geometry obtained by the mirror transformation $\tau\leftrightarrow \rho$, i.e. T-duality in the $x_1$ direction.
Here one obtains the following asymmetric rotation:
\bea
\tau\leftrightarrow \rho\, :\qquad &{~}&X_L\leftrightarrow \tilde X_L=X_L\, ,\nonumber\\
&{~}&X_R\leftrightarrow \tilde X_R=-\bar X_R
\, .
\eea
The action of the $SL(2,{\mathbb Z})_\tau$ transformations now become asymmetric rotations on the dual coordinates,
\bea\label{asymrot1}
\tau\rightarrow -1/\tau\, :\qquad &{~}&\tilde X_L\rightarrow e^{i\theta}\tilde X_L\, ,\quad \theta=-\pi/ 2\, ,\nonumber\\
&{~}&\tilde X_R\rightarrow e^{-i\theta}\tilde X_R\, .\nonumber \\
\eea
where the $SL(2,{\mathbb Z})_\rho$ transformations act symmetrically:
\bea\label{symrot1}
\rho\rightarrow -1/\rho\, :\qquad &{~}&\tilde X_L\rightarrow e^{i\theta}\tilde X_L\, ,\quad \theta=-\pi/ 2\, ,\nonumber\\
&{~}&\tilde X_R\rightarrow e^{i\theta}\tilde X_R\, .
\eea

Although already  basically well known, all these transformation properties of the left- and right-moving coordinates   will turn
out be be important for the closed string non-commutativity. 
But before we analyze  in more detail how non-commutativity can arise, let us make the following general  observations about closed string non-commutativity.
Suppose one finds in some background  the following commutation relations between left- and right-moving complex string coordinates:
\begin{equation}\label{comma}
[X_L,\bar X_L]=-[X_R,\bar X_R]=\Theta\, ,\qquad [X_L, X_R]=[X_L,\bar X_R]=0\, .
\end{equation}
As we will show in section (3.2) for the
case of the shifts in $\sigma$, these commutators correspond to the symmetric rotations of the form eq.(\ref{symrot}), which are induced by the $SL(2,{\mathbb Z})_\tau$
modular transformations. 
They imply the following commutation relation between the left- and right-moving real coordinates:
\begin{equation}
[X_L^1, X_L^2]=-[X_R^1, X_R^2]=i\Theta /2\, .
\end{equation}
Finally going to the position space coordinates $X^{1,2}=X_L^{1,2}+X_R^{1,2}$, one then obtains
\begin{equation}\label{comm}
[X^1, X^2]=[X_L^1+X_R^1, X_L^2+X_R^2]=0.
\end{equation}
So in position space this background corresponds to a commutative geometry.

Now we switch to the dual coordinates with respect to the $x_1$ direction, which is equivalent to perform a T-duality transformation in
the $x_1$ direction. As it is well known, for type II backgrounds, this T-duality transformation exchanges the type IIA string with the type IIB string and
vice versa.
As discussed this acts as $X_R^1\rightarrow -X_R^1$.
Hence the commutator eq.(\ref{comm}) is replaced  by the following expression
\begin{equation}
[\tilde X^1, X^2]=[X_L^1-X_R^1, X_L^2+X_R^2]=i\Theta\, .
\end{equation}
So we see that the dual space coordinates, or equivalently the T-dual background, are now non-commutative.
This justifies our claim in the introduction. 
In the next chapter we will demonstrate this behavior for an explicit example.

In fact, for the
case of the shifts in $\sigma$, the non-commutative coordinates of the T-dual space are precisely associated to the asymmetric rotations eq.(\ref{asymrot})  induced by the duality
transformation $SL(2,Z)_\rho$. They will induce to the following commutation relations between the left- and right-moving complex coordinates:
\begin{equation}\label{commg}
[X_L,\bar X_L]=[X_R,\bar X_R]=\Theta\, ,\qquad [X_L, X_R]=[X_L,\bar X_R]=0\, .
\end{equation}
It now follows that 
\begin{equation}
[X^1, X^2]=[X_L^1+X_R^1, X_L^2+X_R^2]=i\Theta\, .
\end{equation}

In case of constant metric and constant B-field one can easily show that all coordinates and also all dual coordinates are fully commuting.
Therefore we have to pose the question, how is it possible to make the $X$-space (or the $\tilde X$-space) non-commutative by turning on certain
background fields, in analogy to the non-commutative open string space when turning on
a B-field background on the open string D-branes? For that  we need also closed string
states with kind of mixed boundary conditions. In other words, we will need twisted closed string sectors, in analogy to
those open strings, which are stretching between two D-branes with different ${\cal F}$-field backgrounds.
The necessary twisting will be achieved by compactifying the closed string on a twisted torus with geometric fluxes turned on.
These geometric fluxes will still lead to commutative $X$-space, in case of shifts in $\sigma$ (see section 3.2). However for shifts in $\tau$ the geometric fluxes
will lead to a non-commuatative coordinate space (see section 3.3).
After T-duality the geometric fluxes will
be dualized into a non-constant $B$-field background, i.e. into a non-trivial $H$-field. Then, as we will discuss, the dual geometry with non-trivial
$H$-field will be non-commutative for the $\sigma$ shifts.
As emphasized already,  all this is most naturally discussed in the context of T-folds
 resp. of doubled geometry. On this space the T-duality group $SO(d,d)$ has a very natural action.
More generally, 
under T-duality the $H$-field background turns into various  geometries like Taub-Nut spaces or Eguchi Hanson instantons.

\section{The twisted torus and the T-dual $H$-field geo\-metry}

\subsection{Monodromies and duality transformations}

Let us demonstrate the closed string non-commutativity and the associated T-duality transformation rules
by an explicit example, namely
that of a twisted $T^3$ with geometrical fluxes, which is dual to a $T^3$ with $H$-flux. These backgrounds are closely related to
Scherk-Schwarz compactifications and freely acting orbifold compactifications, discussed e.g. in
\cite{Scherk:1978ta,Ferrara:1987es,Ferrara:1987jq,Kounnas:1989dk,Kiritsis:1997ca,Gregori:1999ns,Serone:2003sv}.
Note that this background does not necessary satisfy the supersymmetry conditions.
This is not a problem, however, as we only use this as an illustrative example and one could e.g. fiber this $T^3$ over something else to get a good string background.

To start, take $(x^1,x^2,x^3)$ as the coordinates on the $T^3$. 
First, we consider the twisted torus without $B$-field background. (For superstrings, this would be the relevant background in type IIA). Its metric has the form:
\begin{equation}
{\rm d}s^2={1\over\Im\tau}|{\rm d}x^1+\tau(z){\rm d}x^2|^2+({\rm d}x^3)^2\, ,\qquad B=H=0\, .
\end{equation}
In general, the complex structure $\tau$ of the $T^2$ is not a constant, but it is rather a non-trivial function of circle coordinate $x^3$: 
$\tau(x^3)=f(x^3)$.
One can easily picture this space as a $T^2$ in the $(x^1,x^2)$ directions fibered over an $S^1$ in the $x^3$ direction. As one goes around the $S^1$ base (we assume that the $S^1$ has unit radius), 
\begin{equation}
x^3\rightarrow x^3+2\pi\, ,
\end{equation}
the fiber $T^2$ undergoes a transformation
\begin{equation}
\tau(x^3)\rightarrow\tau(x^3+2\pi)\, .
\end{equation}
If we want to end up with an equivalent fiber after going around the $S^1$, we need to ensure that  this is an $SL(2,{{\mathbb Z}})_\tau$ transformation, i.e.
\begin{equation}
\tau(x^3+2\pi)={a\tau(x^3)+b\over c\tau(x^3)+d}\, .
\end{equation}
These transformation define the monodromy properties of the torus fibration.

Now we perform a T-duality transformation on the $x^1$ direction: this  yields the background of a rectangular $T^3$
with non-trivial $B$- and $H$-field. (In type II compactifications, one now goes from IIA  to IIB  at the same time.) Specifically, this background is simply obtained by replacing the complex structure $\tau(x^3)$
of the twisted torus by the K\"ahler parameter $\rho(x^3)$ of the dual geometry, i.e. $\rho(x^3)=f(x^3)$.
This background is still a non-trivial fibration, since the volume of the $T^2$ as well as the $B$-field\footnote{With this
choice of $B$-field we have picked a gauge where $B_{x^1x^2}(x^3)\neq 0$. In addition, as we will see later, the $H$-field
is quantized such that the following condition is satisfied: $\int_{T^3} H = N$.}
now vary over the $S^1$:
\begin{equation}
{\rm Vol}^{T^2}(x^3)=\Im\rho(x^3)\, ,\qquad B(x^3)=\Re\rho(x^3)\, ,\qquad H(x^3)={d\over dx^3}B(x^3)\, .
\end{equation}
Again we have to require that the fibration is such that going around the $S^1$ circle one changes the K\"ahler parameter $\rho$
only by a target space T-duality transformation, i.e..
\begin{equation}
\rho(x^3+2\pi)={a\rho(x^3)+b\over c\rho(x^3)+d}\, .
\end{equation}
In general, these backgrounds are not anymore geometric manifolds, since global diffeomorphisms 
only close up to T-duality transformations. Therefore they are called T-folds \cite{Hull:2006va}.

Let us now discuss a few examples of monodromies that imply certain choices for the function $f(x^3)$. This discussion is of course
equally valid for the twisted torus and the T-dual  $H$-field background.

\vskip0.2cm
\noindent {\sl (i) Trivial monodromy:}

\vskip0.2cm
\noindent
Here $f(x^3+2\pi)=f(x^3)$. The simplest choice obviously is $f(x^3)={\rm const.}$, but also other choices of period functions are possible.
For $f(x^3)={\rm const.}$, the $H$-field of course vanishes. 

\vskip0.2cm
\noindent {\sl (ii) Parabolic monodromies:}

\vskip0.2cm
\noindent
Parabolic monodromies are basically generated by discrete shift in the function $f(x^3)$:
\begin{equation}
f(x^3)={1\over 2\pi}Hx^3+{\rm const.}\, ,\qquad H\in {\mathbb Z}
\end{equation}
These monodromies are of infinite order.  
On the (IIA) twisted torus (here the so-called nilmanifold), the shift $x^3\rightarrow x^3+2\pi$ has to be followed by a shift in the coordinate $x^1$:
\begin{equation}
x^1\rightarrow x^1+Hx^2\, ,\qquad x^2\rightarrow x^2\, .
\end{equation}
In the dual (IIB) geometry on the flat 3-torus, 
going around the circle $S^1$ means that we shift the linear $B$-field, 
\begin{equation}
B={1\over 2\pi}H\,x^3\, ,
\end{equation}
by an integer: $B\rightarrow B+H$. The $H$-field is constant and is given by an integer number.

\vskip0.2cm
\noindent {\sl (iii) Elliptic monodromies:}

\vskip0.2cm
\noindent
Elliptic monodromies are of finite order. They act as 
${\mathbb Z}_N$-transformations on the $T^2$ coordinates. Example are the order two transformation $f(x^3+2\pi)=-1/f(x^3)$
and the order three transformation $f(x^3+2\pi)=-1/(f(x^3)+1)$.

Let us analyze the first transformation in more detail. For the (IIA) twisted torus, $\tau(x^3)\rightarrow -1/\tau(x^3)$ acts on the 
torus coordinates $x^1$ and $x^2$ as a ${\mathbb Z}_4$ rotation: 
\begin{equation}
x^1\rightarrow x^2\, ,\qquad x^2\rightarrow -x^1\, ,
\end{equation}
or written in complex coordinates $x=x^1+ix^2$:
\begin{equation}
x\rightarrow e^{-2\pi iH}x\, ,\qquad H\in{1\over 4}+{\mathbb Z}\, .
\end{equation}
A function $\tau(x^3)$ with the required monodromy properties is e.g. given as \cite{Hull:2005hk}:
\begin{equation}
\tau(x^3)={(1+i)\cos (Hx^3)+\sin (Hx^3)\over\cos (Hx^3)-(1+i)\sin (Hx^3)}
\, .
\end{equation}
After T-duality (to IIB), we exchange $\tau(x^3)$ with $\rho(x^3)$, and  the $B$-field is given as
\begin{equation}
B(x^3)=\Re\rho(x^3)={\sin(2Hx^3)-2\cos(2Hx^3)\over 2\sin(2Hx^3)+\cos(2Hx^3)-3}\, ,
\end{equation}
and the $H$-field has the form
\begin{equation}
H(x^3)=H{10-12\sin(2Hx^3)-6\cos(2Hx^3)\over (2\sin(2Hx^3)+\cos(2Hx^3)-3)^2}\, .
\end{equation}
This expression can be expanded for small $x^3$ (or  for small $H$), where the $H$-field becomes constant: $H(x^3)=H$.
Note that $H(x^3)$ is a periodic function in $x^3$ with period $4\pi$, 
\begin{equation}
H(x^3+4\pi)=H(x^3)\, ,
\end{equation}
i.e. going twice around the circle does not change the $H$-field.
Going once the circle one gets
\begin{equation}
H(x^3+2\pi)=-H(x^3)\, ,
\end{equation}
These transformation properties of the $H$-field just correspond to the ${\mathbb Z}_4$ order of the considered monodromy transformation.

In addition, also the volume of $T^2$ is varying over circle in the $x^3$-direction:
\begin{equation}
{\rm Vol}^{T^2}(x^3)=\Im\rho(x^3)={1\over1+\sin^2(Hx^3)-\sin(2Hx^3)}\, .
\end{equation}
In fact, going once around the circle exchanges a large $T^2$ with a small $T^2$, which are however equivalent from the closed string point of
view because of  T-duality in the $x^1$- and $x^2$-directions. Therefore this background is not a geometrical background in the usual sense, but a T-fold.
A similar analysis can be also done for the elliptic monodromy of order three.

\subsection{The $\sigma$-monodromies and winding non-commutativity}

\vskip0.2cm
So far we have discussed the monodromy properties of the twisted torus geometry and its dual $B$-field background geometry. 
But now we have to remember that we are considering a closed string moving in this background.
Therefore  we have to consider the closed string boundary conditions in this background.
Let us start with the twisted torus geometry. In the circle $x^3$-direction, the closed string obeys the standard periodicity conditions
\begin{equation}
X^3(\tau,\sigma+2\pi)=X^3(\tau,\sigma)+2\pi N_3\, ,
\end{equation}
where $N_3$ is the winding number of the closed string in the $x^3$-direction.
According to our previous discussion, due to the non-trivial fibration structure the non-trivial boundary conditions
in the $x^3$ direction will also act on the coordinates of the 2-torus, namely on $\tau(x^3)$ resp. on $\rho(x^3)$, in a non-trivial way.
As a consequence, depending on which monodromy properties we are considering, one gets shifted or twisted closed string boundary conditions
in the $X^{1,2}$-directions. Specifically, neglecting the momentum and winding zero modes in these two directions, we obtain
the following closed string boundary conditions for the three monodromies considered before:

\vskip0.2cm
\noindent {\sl (i) Trivial monodromy:}

\vskip0.2cm
\noindent
\begin{equation}
X^{1,2}(\tau,\sigma+2\pi)=X^{1,2}(\tau,\sigma)\, .
\end{equation}

\vskip0.2cm
\noindent {\sl (ii) Parabolic monodromies:}

\vskip0.2cm
\noindent
Now the closed string boundary conditions include the following shift in the coordinate $x^1$:
\begin{equation}
X^1(\tau,\sigma+2\pi)=X^1(\tau,\sigma)+N_3HX^2(\tau,\sigma)\, ,\qquad X^2(\tau,\sigma)=X^2(\tau,\sigma)\, .
\end{equation}

\vskip0.2cm
\noindent {\sl (iii) Elliptic monodromies:}

\vskip0.2cm
\noindent
Here, the transformation $\sigma\rightarrow \sigma+2\pi$ acts as a rotation in the $X^{1,2}$-directions. Namely one obtains
($X(\tau,\sigma)=X^1(\tau,\sigma)+iX^2(\tau,\sigma)$):
\begin{equation}\label{twisted}
X(\tau,\sigma+2\pi)=e^{-2\pi iN_3H}X(\tau,\sigma)\, .
\end{equation}
Note that in both cases the shift parameter or resp. the rotation angle is given by the product $N_3H$, i.e. non-trivial boundary conditions require
a non-zero winding number $N_3$ in connection with a non-vanishing $H$-field.

As a specific example, we now want to 
analyze in more detail the elliptic monodromy corresponding to the
modular transformation $\tau(x^3)\rightarrow -1/\tau(x^3)$; it leads to   the ${\bf Z}_4$ twisted boundary conditions displayed in eq.(\ref{twisted})
 with $N_3H\in {1/4}+{\mathbb Z}$.
 Switching to complex, left- and right-moving closed string  coordinates $X_L$ and $X_R$, we are dealing with the following left-right symmetric
 closed string boundary conditions for the twisted torus geometry:
 \bea\label{symbound}
X_L(\tau,\sigma+2\pi)&=& e^{i\theta}X_L(\tau,\sigma)\, ,\quad \theta=-2\pi N_3H\, ,\nonumber\\
X_R(\tau,\sigma+2\pi)&=& e^{i\theta}X_R(\tau,\sigma)\, .
\eea
The complex string coordinates, which precisely possess these  boundary conditions like a symmetric orbifold 
\cite{Dixon:1985jw,Dixon:1986jc,Hamidi:1986vh,Dixon:1986qv}, 
have the following twisted mode expansion (this mode expansion
is quite similar to the mode expansion of open strings on intersecting D1-branes -- see the appendix):
\bea
X_L(\tau+\sigma)&=&i\sqrt{{\alpha'\over 2}}\sum_{n\in{\mathbb Z}}{1\over n-{\nu}}\alpha_{n-\nu}e^{-i(n-\nu)(\tau+\sigma)}\, ,\qquad\nu={\theta\over2\pi}=-N_3H\, ,\nonumber\\
X_R(\tau-\sigma)&=&i\sqrt{{\alpha'\over 2}}\sum_{n\in {\mathbb Z}}{1\over n+\nu}\tilde\alpha_{n+\nu}e^{-i(n+\nu)(\tau-\sigma)}\, ,
\eea
Then $\bar X_L(\tau,\sigma)$ is given as
\begin{equation}
\bar X_L(\tau+\sigma)=-i\sqrt{{\alpha'\over 2}}\sum_{n\in{\mathbb Z}}{1\over n-{\nu}}\bar\alpha_{n-\nu}e^{i(n-\nu)(\tau+\sigma)}\, ,
\end{equation}
and similarly for $\bar X_R(\tau,\sigma)$.
Here, the $\alpha_{n-\nu}$ and the   $\bar\alpha_{n-\nu}$    are complex, twisted oscillators with the following commutation relations
\begin{equation}
[\alpha_{n-\nu},\bar\alpha_{m-\nu}]=(n-\nu)\delta_{n,m}\, ,
\end{equation}
and likewise for 
$ \tilde\alpha_{n+\nu}$. It follows that $X_L^1(\tau,\sigma)$ has the following form, 
\bea\label{x1}
X_L^1(\tau+\sigma)&=&{X_L(\tau+\sigma)+\bar X_L(\tau+\sigma)\over 2}\nonumber \\
&=&{i\over 2}\sqrt{{\alpha'\over 2}}\sum_{n\in{\mathbb Z}}{1\over n-{\nu}}
\Biggl\lbrack
\alpha^1_{n-\nu}e^{-i(n-\nu)(\tau+\sigma)}   -\alpha^1_{-(n-\nu)}e^{i(n-\nu)(\tau+\sigma)} \nonumber\\
&+&i\biggl(\alpha^2_{n-\nu}e^{-i(n-\nu)(\tau+\sigma)}   +\alpha^2_{-(n-\nu)}e^{i(n-\nu)(\tau+\sigma)}\biggr)\Biggr\rbrack\, ,
\eea
with $\alpha_{n-\nu}=\alpha^1_{n-\nu}+i\alpha^2_{n-\nu}$, and where we have used
that 
\begin{equation}
\alpha_{-(n-\nu)}^*=\bar\alpha_{n-\nu} \, .
\end{equation}
For $\nu=0$, eq.(\ref{x1}) collapses to
\begin{equation}
X_L^1(\tau+\sigma)=
i\sqrt{{\alpha'\over 2}}\sum_{n\neq0}{1\over n}\alpha^1_ne^{-in(\tau+\sigma)}  \, .
\end{equation}
Here we have removed the divergent zero-mode term with $n=0$.
Similar expressions hold for $X_L^2(\tau,\sigma)$ $X_R^1(\tau,\sigma)$ and $X_R^2(\tau,\sigma)$.

Using these mode expansions and the commutation relations among the oscillators we can now compute the commutator between two left- or
right-moving  complex closed string coordinates at equal times $\tau$. One obtains the following result:
\bea
\lbrack X_L(\tau,\sigma),\bar X_L(\tau,\sigma')\rbrack&=&\Theta(\sigma,\sigma'  ) =\alpha'\sum_{n\in{\mathbb Z} } {1\over n-\nu}e^{-i(n-\nu)(\sigma-\sigma')}\, ,\nonumber\\
\lbrack X_R(\tau,\sigma),\bar X_R(\tau,\sigma')\rbrack &=&-\Theta(\sigma,\sigma')\, .
\eea
This series can be written in terms of hypergeometric functions as follows  ($z=e^{i(\sigma'-\sigma)}$):
\begin{equation}
\Theta(\sigma,\sigma')={1\over \nu}z^{\nu}\biggl(~_2F_1(1,-\nu;1-\nu;z)
- ~_2F_1(1,\nu;1+\nu;z^{-1})-1\biggr)\,  .
\end{equation}
For $\sigma=\sigma'$, this sum can be explicitly evaluated:\footnote{Here one can use that $\psi(x)=\partial_x\ln\Gamma(x)$,
$\pi\cot(\pi z)=\psi(1-z)-\psi(z)$, $\psi(1-z)=-\gamma_E-{1\over 1-z}-\sum_{n=1}^\infty \bigl({1\over 1-z+n}-{1\over n}\bigr)$.}
\begin{equation}\label{finalsum}
\Theta=\alpha'\sum_{n\in{\mathbb Z} } {1\over n-\nu}=\alpha'\pi\cot\theta/2=-\alpha'\pi\cot(\pi N_3H)\, .
\end{equation}
Using $H=1/4$, one obtains that $\Theta=-\alpha'\pi$ for winding number $N_3=1$ ($\theta=-\pi/2$), whereas
$\Theta=0$ for $N_3=2$ ($\theta=-\pi$). Also note that $\Theta$ formally diverges for $\theta=0~{\rm mod}~2\pi$. This infinity arises, since the sum eq.(\ref{finalsum}) 
contains the divergent term $1/n$. However for $\theta=0~{\rm mod}~2\pi$ the $1/n$-term has to be excluded from the untwisted mode expansion.
Taking this into account and summing over all terms with $n\neq0$, one indeed obtains $\Theta=0$ for $\theta=0~{\rm mod}~2\pi$; hence $\Theta$ exhibits  a  discontinuity at $\theta=0~{\rm mod}~2\pi$.

We see that the non-vanishing commutators between the left-moving coordinates and the right-moving coordinates have opposite signs.
Therefore we encounter precisely the situation of eq.(\ref{comma}). It follows that the geometric, twisted torus geometry is commutative:
\begin{equation}\label{commx}
[X^1(\tau,\sigma), X^2(\tau,\sigma)]=[X_L^1+X_R^1, X_L^2+X_R^2]=0.
\end{equation}

However switching to the dual coordinates after a T-duality in the  $x^1$-direction one obtains a non-trivial commutator:
\begin{equation}
[\tilde X^1(\tau,\sigma), X^2(\tau,\sigma)]=[X_L^1-X_R^1, X_L^2+X_R^2]=i\Theta\, .
\end{equation}
It means that the dual $H$-field background with varying K\"ahler parameter becomes non-commutative.
This can be explicitly checked by looking at the closed string boundary conditions of the $H$-field background. The closed
string boundary condition $X^3(\tau,\sigma+2\pi)=X^3(  \tau,\sigma)+2\pi N_3$    in the circle direction now corresponds to 
the monodromy transformation
$\rho\rightarrow -1/\rho$ on the 2-torus. 
This transformation implies the following left-right asymmetric boundary conditions, in analogy to the boundary conditions
of an asymmetric orbifold \cite{Narain:1986qm}:
 \bea\label{asymbound}
X_L(\tau,\sigma+2\pi)&=& e^{i\theta}X_L(\tau,\sigma)\, ,\quad \theta=-2\pi N_3H\, ,\nonumber\\
X_R(\tau,\sigma+2\pi)&=& e^{-i\theta}X_R(\tau,\sigma)\, .
\eea
They are consistent with the following mode expansion  (this mode expansion
is quite similar to the mode expansion of open strings on  D2-branes with ${\cal F}$-flux -- see the appendix):
\bea
X_L(\tau+\sigma)&=&i\sqrt{{\alpha'\over 2}}\sum_{n\in{\mathbb Z}}{1\over n-{\nu}}\alpha_{n-\nu}e^{-i(n-\nu)(\tau+\sigma)}\, ,\qquad\nu={\theta\over2\pi}\nonumber\\
X_R(\tau-\sigma)&=&i\sqrt{{\alpha'\over 2}}\sum_{n\in {\mathbb Z}}{1\over n-\nu}\tilde\alpha_{n-\nu}e^{-i(n-\nu)(\tau-\sigma)}\, ,
\eea
Then one 
obtains in the dual space that
\bea
\lbrack X_L(\tau,\sigma),\bar X_L(\tau,\sigma')\rbrack=\lbrack X_R(\tau,\sigma),\bar X_R(\tau,\sigma')\rbrack=
\Theta(\sigma,\sigma'  ) \, ,
\eea
and finally 
\begin{equation}
[ X^1(\tau,\sigma), X^2(\tau,\sigma)]=i\Theta\, .
\end{equation}
So, the T-fold with $H$-flux becomes non-commutative,
the non-commutativity parameter $\Theta$ originates from the winding number $N_3$ in the third direction.

\subsection{The $\tau$-monodromies and momentum non-commutativity}

It is now straightforward to consider those monodromies on the (dual) circle on $x^3$-direction, which correspond to the (not so conventional) boundary conditions with respect to the world sheet coordinate $\tau$ (see also the appendix). Specifically, consider, instead of an winding string in the third direction, a string with momentum number $M_3$.
Then, going through the same kind of arguments and computations as in the previous subsection, one obtains from the "periodicity" condition with
respext to $\tau$,
\begin{equation}
X^3(\tau+2\pi,\sigma)=X^3(\tau,\sigma)+2\pi M_3\, ,
\end{equation}
the following commutators for the twisted torus with geometric fluxes:
\bea
\lbrack X_L(\tau,\sigma),\bar X_L(\tau,\sigma)\rbrack&=&\tilde \Theta =-\alpha'\pi\cot(\pi M_3H)\, ,\nonumber\\
\lbrack X_R(\tau,\sigma),\bar X_R(\tau,\sigma)\rbrack &=&\tilde\Theta\, ,\nonumber\\
\lbrack X_L^1(\tau,\sigma), X_L^2(\tau,\sigma)\rbrack&=&i\tilde \Theta /2\, ,\nonumber\\
\lbrack X_R^1(\tau,\sigma),X_R^2(\tau,\sigma)\rbrack &=&i\tilde\Theta/2\, .
\eea
Hence we now obtain the following non-vanishing commutator between the space coordinates $X^1$ and $X^2$:
\begin{equation}
[X^1(\tau,\sigma), X^2(\tau,\sigma)]=i\tilde\Theta.
\end{equation}

On the other hand,  for the dual geometry, namely for the T-fold with $H$-flux one now gets
\bea
\lbrack X_L(\tau,\sigma),\bar X_L(\tau,\sigma)\rbrack&=&\tilde \Theta =-\alpha'\pi\cot(\pi M_3H)\, ,\nonumber\\
\lbrack X_R(\tau,\sigma),\bar X_R(\tau,\sigma)\rbrack &=&-\tilde\Theta\, ,\nonumber\\
\lbrack X_L^1(\tau,\sigma), X_L^2(\tau,\sigma)\rbrack&=&i\tilde \Theta /2\, ,\nonumber\\
\lbrack X_R^1(\tau,\sigma),X_R^2(\tau,\sigma)\rbrack &=&-i\tilde\Theta/2\, , \nonumber\\
\lbrack X^1(\tau,\sigma), X^2(\tau,\sigma)\rbrack &=&0\, .
\eea
So as before, a non-commutative background is T-dual to a commutative background and vice versa.

\subsection{Combining winding and momentum non-commutativity}

Now consider the general case with non-vanishing momentum and winding numbers in the third direction, i.e. $M_3,N_3\neq0$, which leads to  two
non-trivial monodromies on the (self-dual) circle in $x^3$-direction.
Then combining the results from the previous two subsections, we obtain for the twisted torus with geometrical fluxes the following commutators:
\bea
\lbrack X_L(\tau,\sigma),\bar X_L(\tau,\sigma)\rbrack&=& =-\alpha'\pi\biggl(\cot(\pi M_3H)+\cot(\pi N_3H)\biggr)\, ,\nonumber\\
\lbrack X_R(\tau,\sigma),\bar X_R(\tau,\sigma)\rbrack &=&-\alpha'\pi\biggl(\cot(\pi M_3H)-\cot(\pi N_3H)\biggr)
\, ,\nonumber\\
\lbrack X_L^1(\tau,\sigma), X_L^2(\tau,\sigma)\rbrack&=&-i{\alpha'\pi\over 2}\biggl(\cot(\pi M_3H)+\cot(\pi N_3H)\biggr)
\, ,\nonumber\\
\lbrack X_R^1(\tau,\sigma),X_R^2(\tau,\sigma)\rbrack &=&-i{\alpha'\pi\over 2}\biggl(\cot(\pi M_3H)-\cot(\pi N_3H)\biggr)
\, ,\nonumber \\
\lbrack X^1(\tau,\sigma), X^2(\tau,\sigma)\rbrack &=&-i\alpha'\pi\cot(\pi M_3H)\, .
\eea

Likewise we obtain for the T-fold with $H$-flux:
\bea
\lbrack X_L(\tau,\sigma),\bar X_L(\tau,\sigma)\rbrack&=& =-\alpha'\pi\biggl(\cot(\pi M_3H)+\cot(\pi N_3H)\biggr)\, ,\nonumber\\
\lbrack X_R(\tau,\sigma),\bar X_R(\tau,\sigma)\rbrack &=&-\alpha'\pi\biggl(-\cot(\pi M_3H)+\cot(\pi N_3H)\biggr)
\, ,\nonumber\\
\lbrack X_L^1(\tau,\sigma), X_L^2(\tau,\sigma)\rbrack&=&-i{\alpha'\pi\over 2}\biggl(\cot(\pi M_3H)+\cot(\pi N_3H)\biggr)
\, ,\nonumber\\
\lbrack X_R^1(\tau,\sigma),X_R^2(\tau,\sigma)\rbrack &=&-i{\alpha'\pi\over 2}\biggl(-\cot(\pi M_3H)+\cot(\pi N_3H)\biggr)
\, ,\nonumber \\
\lbrack X^1(\tau,\sigma), X^2(\tau,\sigma)\rbrack &=&-i\alpha'\pi\cot(\pi N_3H)\, .
\eea
So we see that now both backgrounds are non-commutative in $X$-space and also in $\tilde X$-space.
Hence the situation is now symmetric with respect to T-duality.

\section{Non-commutative geometry and (dual) momentum dependent space uncertainty}

So far, the right hand sides of the commutators of $X^1$ and $X^2$ were determined by the winding and momentum numbers $N_3$ and $M_3$.
However these two numbers  are nothing else than the discrete eigenvalues of the canonical momentum operator $P^3$ and of the dual momentum  operator
$\tilde p^3$
of the closed string on the circle
in the third direction.\footnote{This section, which is not in the first version of our paper,
was partly inspired by the recent work \cite{BlPl}, namely to argue that the commutator of two space coordinates is not a c-number
but rather an operator.}
 Therefore we now conjecture that the full non-commutative algebraic structure is obtained by
  replacing the momentum number $M_3$ by the (zero mode) momentum operator $p^3$ and the winding number $N_3$ by the dual
  momentum operator $\tilde p^3$ in all above equations:
  \begin{equation}
  M_3\equiv \sqrt{\alpha'}\,  p^3\, ,\qquad N_3\equiv\sqrt{\alpha'} \, \tilde p^3\, .
  \end{equation}
   Furthermore expanding the $\cot$-function  (around $\pi/4$) with respect to these operators, we obtain at
  linear order in the (dual) momenta the following refined commutation relations, where the right hand sides of the commutators are now given in terms of (dual) momentum operators.
  First for the twisted torus with geometrical fluxes we get ($\alpha'=l_s^2$):
\bea\label{aset}
\lbrack X_L(\tau,\sigma),\bar X_L(\tau,\sigma)\rbrack&=& 2l_s^3\pi^2H(p^3+\tilde p^3)=2l_s^3\pi^2H\, p_L^3\, ,\nonumber\\
\lbrack X_R(\tau,\sigma),\bar X_R(\tau,\sigma)\rbrack &=&2l_s^3\pi^2H(p^3-\tilde p^3)=2l_s^3\pi^2H\, p_R^3
\, ,\nonumber\\
\lbrack X_L^1(\tau,\sigma), X_L^2(\tau,\sigma)\rbrack&=&il_s^3\pi^2H(p^3+\tilde p^3)=il_s^3\pi^2H\, p_L^3
\, ,\nonumber\\
\lbrack X_R^1(\tau,\sigma),X_R^2(\tau,\sigma)\rbrack &=&il_s^3\pi^2H(p^3-\tilde p^3)=il_s^3\pi^2H\, p_R^3
\, ,\nonumber \\
\lbrack X^1(\tau,\sigma), X^2(\tau,\sigma)\rbrack &=&2il_s^3\pi^2H~p^3\, .
\eea
We see that the commutator between two space coordinates is now given in terms of the momentum operator $p^3$.
Here, the constant $H$ on the right hand side just corresponds to the geometrical flux of the twisted torus.
We will further discuss this result and the corresponding uncertainty relations at the end of this section.

Before we come to this discussion, let us determine the analogous commutators for the dual geometry, the T-fold with $H$-flux. Here we get:
\bea\label{bset}
\lbrack X_L(\tau,\sigma),\bar X_L(\tau,\sigma)\rbrack&=& 2l_s^3\pi^2H(p^3+\tilde p^3)=2l_s^3\pi^2H\, p_L^3\, ,\nonumber\\
\lbrack X_R(\tau,\sigma),\bar X_R(\tau,\sigma)\rbrack &=&-2l_s^3\pi^2H(p^3-\tilde p^3)=-2l_s^3\pi^2H\, p_R^3
\, ,\nonumber\\
\lbrack X_L^1(\tau,\sigma), X_L^2(\tau,\sigma)\rbrack&=&il_s^3\pi^2H(p^3+\tilde p^3)=il_s^3\pi^2H\, p_L^3
\, ,\nonumber\\
\lbrack X_R^1(\tau,\sigma),X_R^2(\tau,\sigma)\rbrack &=&-il_s^3\pi^2H(p^3-\tilde p^3)=-il_s^3\pi^2H\, p_R^3
\, ,\nonumber \\
\lbrack X^1(\tau,\sigma), X^2(\tau,\sigma)\rbrack &=&2il_s^3\pi^2H~\tilde p^3\, .
\eea
Now the constant $H$ on the right hand side corresponds to the $H$-flux of the T-fold. 
Note that these two sets of equations precisely have the correct transformation properties under T-duality.

Equations (\ref{aset}) and (\ref{bset}) can be neatly combined for backgrounds that possess both geometrical and $H$-fluxes. Denote the geometrical flux
by $f^{123}$ and the $H$-flux by $H^{123}$, which are T-dual to each other (see eq.(\ref{geomh})), we can write these equations as
\bea\label{cset}
\lbrack X_L(\tau,\sigma),\bar X_L(\tau,\sigma)\rbrack&=& 2l_s^3\pi^2(f^{123}+H^{123})\, p_L^3\, ,\nonumber\\
\lbrack X_R(\tau,\sigma),\bar X_R(\tau,\sigma)\rbrack &=&2l_s^3\pi^2(f^{123}-H^{123})\, p_R^3
\, ,\nonumber\\
\lbrack X_L^1(\tau,\sigma), X_L^2(\tau,\sigma)\rbrack&=&il_s^3\pi^2(f^{123}+H^{123})\, p_L^3
\, ,\nonumber\\
\lbrack X_R^1(\tau,\sigma),X_R^2(\tau,\sigma)\rbrack &=&il_s^3\pi^2(f^{123}-H^{123})\, p_R^3
\, ,\nonumber \\
\lbrack X^1(\tau,\sigma), X^2(\tau,\sigma)\rbrack &=&2il_s^3\pi^2(f^{123}~p^3+H^{123}\, \tilde p^3)\, .
\eea

Now let us discuss the physical meaning of these equations. First note that we have two sets of commutators in doubled geometry,
namely one for the left-movers and another one for the right-movers, and that 
they are fully consistent with T-duality (in our example in $x^1$-direction).
Second, looking at the last equations in (\ref{aset}) or in  (\ref{cset}), we see that in a curved gravitational background with non-trivial geometrical flux,
the commutator of the two closed string coordinates $X^1$ and $X^2$ is proportional to the product of the geometrical flux times the
momentum in the third direction:
\begin{equation}
\lbrack X^1, X^2\rbrack =2il_s^3\pi^2f^{123}\, p^3\, .
\end{equation}
Therefore from this commutation relation we can derive the following uncertainty relation for the coordinates $X^1$ and $X^2$:
\begin{equation}
(\Delta X^1)^2(\Delta X^2)^2\geq \pi^4l_s^6(f^{123})^2\, \, \langle p^3\rangle^2\, .
\end{equation}
This means that the uncertainty in $X^1$ and $X^2$ is proportional to the expectation value of the momentum operator in the third direction.
So, for small momenta, this uncertainty almost vanishes. However for high momenta of the order $\langle p^3\rangle\sim l_s^{-1}$ in the third direction
of the space, the spatial uncertainty in the other two directions is of the order
$l_s^2$. In other words, the position of a string, that moves with high velocity in the third direction, cannot not be precisely determined
in the other two directions.
This is  what one could perhaps expect in a theory of quantum gravity on a curved space: space becomes non-commutative, i.e. fuzzy, at high energies,
but is almost commutative at low energies.  Also note, since $p^3$ and $x^3$ have standard, canonical Heisenberg commutation relation,
\begin{equation}
\lbrack p^3,X^3\rbrack =-i\, ,
\end{equation}
we get
\begin{equation}
\lbrack\lbrack X^1, X^2\rbrack,X^3\rbrack =2\pi^2f^{123} \, l_s^3\, .
\end{equation}
Hence the uncertainty of the product between the commutator $\lbrack X^1, X^2\rbrack$ and $X^3$ is just the usual
Heisenberg uncertainty relation in the third direction, 
and
is given by
\bea
\bigl(\Delta\lbrack X^1, X^2\rbrack\bigr)^2\,(\Delta X^3)^2&=&\pi^4(f^{123})^2 \, l_s^6(\Delta p^3)^2(\Delta X^3)^2\nonumber\\
&\geq&\pi^4(f^{123})^2 \, l_s^6\, .
\eea

Of course, one can derive similar uncertainty relations for $H$-flux backgrounds, where one replaces the momentum $p^3$ by the dual momentum
$\tilde p^3$.

\section{Summary $\&$ Outlook}

Let us summarize this paper with a few remarks, questions and also with a few speculations:

\begin{itemize}

\item
The main result of this paper is that we gave some evidence that a  closed string on a  curved space (3-dimensional torus) with a non-trivial
geometrical flux $f^{123}$ and  $H$-field $H^{123}$ sees
a non-commu\-ta\-tive geometry  along two  directions, which is determined by the (dual) momentum in the remaining third direction:
\begin{equation}
\lbrack X^1(\tau,\sigma),X^2(\tau,\sigma)\rbrack =       2il_s^3\pi^2(f^{123}\, p^3+H^{123}\, \tilde p^3)\, .
\end{equation}
In a more symmetric situation between all three coordinates, we can
combine this non-commutative closed string geometry together with the standard Heisenberg algebra of coordinates and momenta, and we
are led to the conjecture that
the following interesting algebraic structure (here in three dimensions) is underlying closed string non-commutative geometry:
\begin{equation}\label{newalgebra}
\lbrack x^i,x^j\rbrack \sim
ic\, \epsilon^{ijk}\, p^k\, ,\qquad \lbrack p^i,p^j\rbrack \sim
ic'\, \epsilon^{ijk}\, p^k\, ,\qquad \lbrack x^i,p^j\rbrack=i\, \delta^{ij}\,,
\end{equation}
 where $c$ is a constant proportional to $l_s^3$, and
 $c'$ is a constant proportional to $l_s^{-1}$,

\item
We made several assumptions in our computation; first, the mode expansion of the coordinates we used is basically the one of a free, twisted boson.
Of course the metric and $B$-field background depend on the coordinates. So in principle we are dealing with an interacting $\sigma$-model.
Therefore our ansatz  should be regarded as lowest order approximation in the $H$-field. This seems to be reasonable, because the back reaction of the
$H$-field, namely its contribution to the $\beta$-function  is quadratic in $H$.
Secondly, in our computation we have assumed a particular gauge, in which the $B$-field components are only along the $x^1$- and $x^2$-directions
of the 3-dimensional torus; in addition the $B$-field  only depends on the third coordinate $x^3$.

\item

Studying the $SU(2)$ WZW  model with a quantized $H$-field, the authors of \cite{BlPl} (see also \cite{Cornalba:2001sm}) give some evidence that the corresponding closed string 
 geometry is not only non-commutative, but 
is even non-associative. 
The basic object, studied in \cite{BlPl}, is the 3-bracket
\begin{equation}
\lbrack\lbrack X^1, X^2\rbrack,X^3\rbrack +{\rm cyclic}\, .
\end{equation}
We can confirm the non-associativity of \cite{BlPl}, namely that for the two dual backgrounds, we studied in the paper,
the 3-bracket is indeed non-vanishing:
\begin{equation}
\lbrack\lbrack X^1, X^2\rbrack,X^3\rbrack +{\rm cyclic}=\pi^2f^{123} \,l_s^3\, .
\end{equation}
In fact, this non-associativity is a general property of the algebra displayed in eq.(\ref{newalgebra}), which always implies that
\begin{equation}
\lbrack\lbrack x^i, x^j\rbrack,x^k\rbrack +{\rm cyclic}=3\,c\,\epsilon^{ijk}\, .
\end{equation}
These issues and their interpretations will be further investigated in \cite{BDLP}.

So, whereas open string non-commutative geometry is associative and is related to associative open string field theory \cite{Witten:1985cc}, the closed string geometry could be also related to the  non-associativity found in closed string field theory 
\cite{Strominger:1987ad,Zwiebach:1992ie},
as well as in
the more recent doubled closed string field theory of 
\cite{Hull:2009mi,Hull:2009zb,Hohm:2010jy,Hohm:2010pp}.

\item
Note that T-dual geometry of the twisted torus
is actually not a geometrical space in the usual sense, it  is rather a T-fold, since the transition functions of global diffeomorphisms
(in our example $x^3\rightarrow x^3+2\pi$) are  not given by   usual $SO(d)$ transformations but by more general duality transformations  in $SO(d,d;{\mathbb Z})$
(in our example $\rho\rightarrow1/\rho$).  
May be that the emergence of the non-commutative geometry
is also related to the fact that we are dealing with  non-geometrical compactifications (the T-fold).

\item
In  a given duality frame, we can always express the non-commutative geometry by non-vanishing commutation relations
between some space  coordinates $X^i$ and some other dual space coordinates $\tilde X^j$,:
\begin{equation}
\lbrack X^i,\tilde X^j\rbrack =i\Theta^{ij}\sim i p^k\, .
\end{equation}
In fact,
the simultaneous existence of coordinates and dual coordinates, i.e. the introduction  of doubled geometry  is essential for the emergence of closed string non-commutative
geometry in the form we have discussed it in this paper.

\item
What about closed string non-commutative gravity?  
Recall, for open strings on D-branes, the non-commutativity is first determined by the closed $B$-field background. In addition,
the open gauge $F$-field background  also contributes to the open string non-commutativity on the D-branes. 
The  non-commutative gauge theory \cite{seiwit,Madore:2000en,Jurco:2000dx,Jurco:2001rq} on the D-branes with $F$-field background has a mathematically very interesting structure, it is 
determined by the Moyal-Weyl star-product. Via the the Seiberg-Witten (SW) map \cite{seiwit}, the non-commutative gauge theory can be  mapped to 
a commutative gauge theory, similar to the T-duality between the commutative torus (D-branes at angles) and the non-commutative torus  
(D-branes with F-flux). In fact, the  (Abelian) gauge theory can be described by the non-linear Born-Infeld action, which contains
an infinite number of higher order term in $F$. 
Could something analogous  be also true for non-commutative closed string geometry,  could there be an analogous SW map,
which maps a non-commutative (and perhaps non associative) gravity theory to a commutative gravity theory?
For that to work, the closed string non-commutativity should not only depend on the $H$-field, but for curved spaces also on the curvature of
the space geometry itself (in analogy to the $F$-field dependence of the open string
non-commutativity). In this case a non-commutative version of gravity along the lines of \cite{Aschieri:2005yw} could be perhaps  formulated.
Via T-duality resp. via the closed string SW map, non-commutative gravity could be possibly mapped to a gravity theory
on a commutative space, which then contains an infinite number of higher order curvature terms (in analogy to the gauge Born-Infeld action).

\item
In general,
six-dimensional, supersymmetric $H$-flux backgrounds as well
as the non-geometric  backgrounds can be best described in terms of so-called
generalized geometry, which generalizes the $SU(3)$ group structures of the geometric
spaces to background with  generalized $SU(3)\times SU(3)$ group structures 
\cite{Grana:2004bg,Jeschek:2005ek,Jeschek:2004wy,Grana:2005sn,Grana:2006is,Grana:2006hr,Gmeiner:2007jn,Koerber:2007xk,Grana:2008yw,Lust:2008zd}.
In analogy to doubled geometry, 
in case of generalized group structure backgrounds one considers the direct sum of the tangent space and the co-tangent space: $T\oplus T^*$.
It follows that the $SU(3,3)$ duality group has a natural action on spaces with generalized group structure.
T-duality is  replaced by mirror symmetry, and the symplectic structure  is exchanged with the complex structure and vice versa
under T-duality.
So it would be interesting to see if non-commutative geometry arises for  $SU(3)\times SU(3)$ group structure backgrounds in a natural way
(see also 
\cite{Kapustin:2003sg}).

\item
Finally note that the string picture, where the  closed string space geometry  cannot be dissolved beyond the string scale due to T-duality and closed
string non-commutativity, is complementary to the recent work
of Dvali and Gomez \cite{Dvali:2010bf}. They argue that in pure Einstein gravity the production of black holes sets the shortest possible dissolvable
distance, which is just  the Planck length.

\end{itemize}

\section*{Acknowledgements}

This work is supported by the Munich Excellence Cluster for Fundamental Physics "Origin and Structure of the Universe" 
as well as by funding from "LMUexcellent".
I like to thank R. Blumenhagen, A. Deser, E. Kiritsis, C. Kounnas, E. Plauschinn, S. Stieberger and D. Tsimpis for useful discussions.
Moreover I am indebted to R. Blumenhagen and E. Plauschinn for informing me about their paper \cite{BlPl} prior to publication.
Their results about closed string non-commutativity and non-associativity in WZW models with $H$-field encouraged me to write up this paper,  which
I thought about already for several years.

\section*{Appendix}

In this appendix we like to  argue that closed string momentum and winding modes
 define in some sense D-branes in closed string theory.
The basic idea is to  compare the closed string periodicity conditions  with open string boundary conditions at the two ends of the open string.

\vskip0.3cm

\noindent
{\sl Open strings:}

\vskip0.3cm

Let us first consider the open string and 
 briefly recall how non-commutativity arises for open strings
in a non-trivial ${\cal F}$-field background.
First  consider an open string $X(\tau,\sigma)$ in one spatial direction.
It has Neumann (N) boundary conditions if at its ends $\sigma=0,\pi$ the open string coordinate obeys:
\begin{equation}
N:\qquad \partial _\sigma X=0\, .\label{neum}
\end{equation}
On the other hand, Dirichlet (D)  boundary conditions mean that at $\sigma=0,\pi$ the open string has the following boundary conditions:
\begin{equation}
D:\qquad \partial _\tau X=0\, .\label{dirich}
\end{equation}

Following the notation of \cite{Blumenhagen:2000fp},
we next consider a 2-dimensional torus with coordinates $X_1$ and $X_2$, where 
we start 
with two D1-branes intersecting at an arbitrary angle $\theta_2-\theta_1$.
The open string boundary condition for a strings that ends on the first D1-brane  are at $\sigma=0$
\bea
\label{sigmanull}N:\qquad\partial_\sigma X_1 + \tan\theta_1\, 
                   \partial_\sigma X_2 &=&0, \nonumber \\
                   D:\qquad   \partial_\tau X_2 - \tan\theta_1\, \partial_\tau X_1
                      &=&0.
                      \eea
                      Hence for $\theta_1=0$ the open string has Neumann boundary conditions in the $X_1$ direction
                      and Dirichlet boundary conditions in $X_2$, i.e. the D1-brane is along the $X_1$ axis. For $\theta_1=\pi/2$ the situation
                      is reversed.
                      
At $\sigma=\pi$  we either require the same boundary conditions
               in case the string also ends the first D1-brane.
   However if at $\sigma=\pi$ the open string ends at the second D1-brane, we have to require         
       \bea\label{sigmanullb} N:\qquad\partial_\sigma X_1 + \tan\theta_2\, 
                   \partial_\sigma X_2, &=&0 ,\nonumber\\
                      D:\qquad\partial_\tau X_2 - \tan\theta_2\, \partial_\tau X_1
                      &=&0\, ,
                      \eea     
                      Introducing complex coordinates, $X=X^1+iX^2$, the D1 boundary conditions for the branes at angles can be also
                      written as (at $\sigma=0,\pi$)
       \bea\label{sigmanullf} N:\qquad e^{-\theta_i}\partial_\sigma X + e^{i\theta_i}\, 
                   \partial_\sigma \bar X, &=&0 ,\nonumber\\
                      D:\qquad\
                      e^{-\theta_i}\partial_\tau X - e^{i\theta_i}\, 
                   \partial_\tau \bar X, &=&0
                       \, .
                      \eea

The mode expansion satisfying these two boundary conditions (\ref{sigmanull}) and (\ref{sigmanullb}) 
for an open string with its ends on the two different branes
contains twisted oscillators $\alpha_{n+\nu}$ and  looks like
\bea\label{modeg}  X_1=x_1 &+&
    i \sqrt{\alpha'} \sum_{n\in Z}  {\alpha_{n+\nu}\over n+\nu} 
           e^{-i(n+\nu) \tau} \cos[(n+\nu)\sigma +\theta_1] + \nonumber\\
    &{~}&i \sqrt{\alpha'} \sum_{m\in Z}  {\alpha_{m-\nu}\over m-\nu} 
           e^{-i(m-\nu) \tau} \cos[(m-\nu)\sigma -\theta_1], \nonumber\\
    X_2=x_2 &+&
    i \sqrt{\alpha'} \sum_{n\in Z}  {\alpha_{n+\nu}\over n+\nu} 
           e^{-i(n+\nu) \tau} \sin[(n+\nu)\sigma +\theta_1] -\nonumber\\
    &{~}&i \sqrt{\alpha'} \sum_{m\in Z}  {\alpha_{m-\nu}\over m-\nu} 
           e^{-i(m-\nu) \tau} \sin[(m-\nu)\sigma -\theta_1], 
           \eea
with $\nu=(\theta_2-\theta_1)/\pi$. (Note that for $\nu\neq0$ this mode expansion is analogous
to the mode expansion of a string in a twisted sector of an orbifold.) Using the usual commutation relation 
\begin{equation}
\label{osci}{  [\alpha_{n+\nu}, \alpha_{m-\nu}]=(n+\nu)\, \delta_{m+n,0} }
\end{equation}
and the vanishing of the commutator of the center of mass coordinates $x_1$ 
and $x_2$ one can easily show that for D-branes at angles the 
general equal time commutator vanishes
\begin{equation} [ X_i(\tau,\sigma), X_j(\tau,\sigma') ]=0 .
\end{equation}
Therefore, the geometry of D-branes at angles, but without background 
gauge fields, is always commutative.  

Performing  open string T-duality in the $X_1$ direction implies exchanging the world sheet coordinates $\tau$
and $\sigma$ for $X_1$. The two intersecting D1-branes will become two D2-branes, and
the geometrical intersection angles $\theta_1$ and $\theta_2$ of the two D1-branes now correspond to constant gauge field strength background ${\cal F}_{1,2}$-fields for D2-branes on the dual torus (${\cal F}_{1,2}=\cot\theta_{1,2}$).
Hence one gets the two 
mixed boundary conditions for the open strings
\bea\label{sigmanullc}
  \partial_\sigma X_1 + \cot\theta_1\, 
                   \partial_\tau X_2 &=&0 ,\nonumber\\
                      \partial_\sigma X_2 - \cot\theta_1\, \partial_\tau X_1
                      &=&0 
                      \eea
at $\sigma=0$ and 
\bea\label{sigmanulld} 
 \partial_\sigma X_1 + \cot\theta_2\, 
                   \partial_\tau X_2 &=&0, \nonumber\\
                      \partial_\sigma X_2 - \cot\theta_2\, \partial_\tau X_1
                      &=&0 
                      \eea
at $\sigma=\pi$.
Again it is easy to write these mixed boundary conditions in terms of the complex coordinate $X=X^1+iX^2$.
The mode expansion satisfying these boundary conditions is
\bea\label{modeh} X_1=x_1& -&
    \sqrt{\alpha'} \sum_{n\in Z}  {\alpha_{n+\nu}\over n+\nu} 
           e^{-i(n+\nu) \tau} \sin[(n+\nu)\sigma +\theta_1] - \nonumber\\
    &{~}&\sqrt{\alpha'} \sum_{m\in Z}  {\alpha_{m-\nu}\over m-\nu} 
           e^{-i(m-\nu) \tau} \sin[(m-\nu)\sigma -\theta_1], \nonumber\\
    X_2=x_2 &+&
    i \sqrt{\alpha'} \sum_{n\in Z}  {\alpha_{n+\nu}\over n+\nu} 
           e^{-i(n+\nu) \tau} \sin[(n+\nu)\sigma +\theta_1] -\nonumber\\
    &{~}&i \sqrt{\alpha'} \sum_{m\in Z}  {\alpha_{m-\nu}\over m-\nu} 
           e^{-i(m-\nu) \tau} \sin[(m-\nu)\sigma -\theta_1].
           \eea
Now one can compute the commutator
\begin{equation}\label{commd}  [ X_1(\tau,\sigma), X_2(\tau,\sigma') ]=[x_1,x_2]+
                   i\sum_{n\in Z} {2\alpha' \over n+\nu} \sin[(n+\nu)\sigma
                   +\theta_1] \sin[(n+\nu)\sigma' +\theta_1] .
                   \end{equation}
 As has been shown in \cite{chen,chu,Blumenhagen:2000fp}, for $\sigma=\sigma'=0$
the evaluation of the sum in (\ref{commd}) yields
\begin{equation}\label{comme}  [ X_1(\tau,0), X_2(\tau,0) ]=-{2\pi i \alpha' {\cal F}_1\over 
      1+{\cal F}_1^2}
      \end{equation}
and for $\sigma=\sigma'=\pi$ one analogously obtains
\begin{equation}\label{commf}  [ X_1(\tau,\pi), X_2(\tau,\pi) ]={2 \pi i\alpha' {\cal F}_2\over 
1+{\cal F}_2^2}.
\end{equation}
Thus, the  coordinates only non-commute at the boundary of the word-sheet, 
where the commutator can be expressed entirely in terms of the gauge field 
on the local D-brane. 
Only for
$\theta_1=\theta_2\in\{0,\pi/2\}$ the $X_1$ and $X_2$ coordinates commute on the
D-brane, in all other cases the end points see a non-commuting
space-time. 
Moreover, in the compact case for rational D-branes
the non-commutative theory on the D-branes
is mapped via T-duality to a commutative theory on D-branes at angles. 
This is only a special example
of the more general rule pointed out in \cite{seiwit} 
that for rational points the non-commutative torus is T-dual to a commutative 
one.

\vskip0.3cm

\noindent
{\sl Closed strings:}

\vskip0.3cm

First consider a closed string on a circle, which is a pure momentum state, i.e. has vanishing winding number $N=0$.
The momentum states propagate in $X$-space; they
obey the standard  periodicity conditions along the closed string in $\sigma$-direction.
We regard these momentum string boundary conditions
 to be analogous to the open string Neumann
boundary conditions eq.(\ref{neum}) and hence call them closed string "Neumann" conditions in $X$-space:
\bea
"N":\qquad 
X(\tau+2\pi,\sigma)&=&X(\tau,\sigma)+2\pi M/R\nonumber\\
X(\tau,\sigma+2\pi)&=&X(\tau,\sigma)\, .
\eea
On the other hand, a momentum state does not propagate in the dual space; hence it possesses "Dirichlet" boundary conditions 
in the $\tilde X$-direction:
\bea
"D":\qquad \tilde X(\tau+2\pi, \sigma)&=&\tilde X(\tau, \sigma)\, ,\nonumber\\
\tilde X(\tau, \sigma+2\pi)&=&\tilde X(\tau, \sigma)+2\pi M/R
\eea

Second, a pure winding state with vanishing momentum number $M=0$ does not propagate in $X$-space, and it satisfies the following closed string "Dirichlet"
boundary conditions in $X$-space and "Neumann" boundary conditions in $\tilde X$-space:
\bea
"N":\qquad \tilde X(\tau+2\pi,\sigma)&=&\tilde X(\tau,\sigma)+2\pi NR\, ,\nonumber\\
\tilde X(\tau, \sigma+2\pi)&=&\tilde X(\tau, \sigma)\, ,\nonumber\\
"D":\qquad X(\tau+2\pi,\sigma)&=&X(\tau,\sigma)\, ,\nonumber\\
X(\tau,\sigma+2\pi)&=&X(\tau,\sigma)+2\pi NR\, .
\eea

 In view of these closed string boundary conditions,  one can loosely say that the momentum states
propagate on the a kind of "D1-brane"
in the doubled $(X,\tilde X)$-space, where the "D1-brane" is along the $X$-direction.
On the other hand, a pure winding state satisfied "Neumann" boundary conditions along $\tilde X$. Therefore winding states
live  on a kind of "D1-brane" along $\tilde X$.
 T-duality exchanges a "D1-brane" in $X$-direction with a "D1-brane" in $\tilde X$-direction.\footnote{Of course, it would be amusing to
 speculate that in doubled gravity these "D-branes" arise as dynamical solutions.}

We can also  extend the discussion about the closed string "D-branes" and their boundary conditions
for the toroidal compactifications. 
To be specific, we consider the closed string boundary of the twisted torus compactification discussed before.
The boundary conditions with respect to $\sigma\rightarrow \sigma+2\pi$ were already worked out in section 3 (see eq.(\ref{symbound})).
Now we include also the periodicity conditions with respect to $\tau\rightarrow \tau+2\pi$.
Then we obtain for the complex closed string coordinates $X(\tau,\sigma)=X_L(\tau+\sigma)+X_R(\tau-\sigma)$  as well as for its dual coordinate (T-duality in $x^1$-direction)
$\tilde X(\tau,\sigma)=\tilde X_L(\tau+\sigma)+\tilde X_R(\tau-\sigma)$ the following set of boundary conditions:
\bea
 "N":\qquad X(\tau,\sigma+2\pi)&=&e^{i\theta} X(\tau,\sigma)\, ,\nonumber\\
 "D":\qquad \tilde X(\tau+2\pi,\sigma)&=&e^{i\theta} \tilde X(\tau,\sigma)\, .
\eea
So we see that the momentum states satisfy rotated (i.e. twisted) "Neumann" boundary conditions in the 2-dimensional $X$-space and
"Dirichlet" boundary conditions in the 2-dimensional $\tilde X$-space.
Of course, for the dual $H$-field background the "Neumann" and "Dirichlet" directions are exchanged:
\bea
 "N":\qquad \tilde X(\tau,\sigma+2\pi)&=&e^{i\theta} \tilde X(\tau,\sigma)\, ,\nonumber\\
 "D":\qquad  X(\tau+2\pi,\sigma)&=&e^{i\theta}  X(\tau,\sigma)\, .
\eea

\end{document}